%Paper: hep-th/9309049
%From: ruiz@nikhef.nl (fernando ruiz)
%Date: Thu, 9 Sep 1993 11:27:14 +0200

%
%
%
\input phyzzx.tex
\tolerance=1000

% Some definitions
%
% This is to number aligned eqs. independently
%

%
% This is to have the acknowledgmentes and references headings in
% lower case
%

%
\def\refoutlw{\par\penalty-400\vskip\chapterskip
   \spacecheck\referenceminspace
   \ifreferenceopen \Closeout\referencewrite \referenceopenfalse \fi
   \line{\fourteenbf\hfil References\hfil}\vskip\headskip
   \input \jobname.refs
   }
%
% Caligraphic characters
%

%
% Greek letters
%
\def\a{\alpha}

\def\d{\delta}
\def\ga{\gamma}
\def\la{\lambda}
\def\m{\mu}
\def\n{\nu}
\def\r{\rho}

\def\ee{\epsilon}
\def\gm{\Gamma}

%
%Miscellaneous
%

%

%
\def\RR{{\rm I\!\!\, R}}
\def\ACS{\scriptscriptstyle ACS}
\def\TMA{\scriptscriptstyle TMA}
\def\SL{{\rm SL}}
\def\TW{{\rm TW}}
\def\tu{\tau_1}
\def\td{\tau_2}
\def\idp{\int { d^3 \!p \over (2\pi)^3}\>}
\def\idpD{\int { d^D \!p\over (2\pi)^D}\>}

\def\to{\rightarrow}

%%%%%%%%%%%%%%%%%%%%   References  %%%%%%%%%%%%%%%%%%%%%%%%%

\REF\schwartz{A. Schwartz, Commun. Math. Phys. {\bf 67} (1979) 1.}
\REF\invariant{C. Tze, Int. J. Mod. Phys. {\bf A3} (1988) 1959.
     \nextline
     E. Witten, Commun. Math. Phys. {\bf 121} (1989) 351.}
\REF\gauss{G. C\u alug\u areanu, Rev. Math. Pures Appl. {\bf 4}
     (1959) 5; Czech. Math. J. {\bf 11} (1961) 588.  \nextline
     W.F. Pohl, J. Math. Mech. {\bf 17} (1968)693. \nextline
     J.H. White, Am. J. Math. {\bf 91} (1969) 693.}
\REF\polyakov{A.M. Polyakov, Mod. Phys. Lett. {\bf A3} (1988) 325.}
\REF\writhing{F. Brock Fuller, Proc. Nat. Acad. Sci. USA {\bf 68}
     (1971) 815.}
\REF\deser{S. Deser, R. Jackiw and S. Templeton, Phys. Rev. Lett.
     {\bf 48} (1982) 975. \nextline
     G. Dunne, R. Jackiw and C.A. Trugenberger, Phys. Rev.
     {\bf D41} (1990) 661.}
\REF\TMYM{G. Giavarini, C.P. Martin and F. Ruiz Ruiz, Nucl. Phys.
     {\bf B381} (1992) 222.}
\REF\wpapers{T.H. Hasson, A. Karlhede and M. Ro\u cek, Phys. Lett.
     {\bf B225} (1989) 92. \nextline
     A. Coste and M. Makowka, Nucl. Phys. {\bf B342} (1990) 721.}
\REF\hepp{K. Hepp, {Th\`eorie de la renormalisation}, Lectures Notes
     in Physics, vol. 2 (Springer, Berlin 1969).}
\REF\wrenormalized{A.M. Polyakov, Nucl. Phys. {\bf B164} (1979) 171.
     \nextline
     V.S. Dotsenko and S.N. Vergeles, Nucl. Phys. {\bf B169} (1980)
     527. \nextline
     R.A. Brandt, F. Neri and M. Sato, Phys. Rev. {\bf D24} (1981)
     879. \nextline
     J.L. Gervais and A. Neveu, Nucl. Phys. {\bf B163} (1980) 189.}
\REF\hamiltonian{M. Asorey, F. Falceto and S. Carlip, {\it
     Chern-Simons states and topologically massive gauge theories},
     University of California at Davis and University of Zaragoza
     preprints UCD-93-7 and DFTUZ 93.5 (hep-th/9304081).}
\REF\epstein{H. Epstein and V. Glaser, Ann. Inst. Henri Poincar\'e
     {\bf XIX} (1973) 211.}
\REF\collins{J.C. Collins, {\it Renormalization} (Cambridge
     University Press, Cambridge 1987).}

%%%%%%%%%%%%%%%%%  The paper  %%%%%%%%%%%%%%%%%

\mathsurround=1pt
\rightline{FTUAM 93/27}
\rightline{NIKHEF-H 93/17}
\rightline{UPRF 93/376}
\date{}
\vskip 1.5 true cm

\titlepage

\title{{\seventeenbf Abelian Chern-Simons theory as the strong
        large-mass limit of topologically massive abelian gauge
        theory: the Wilson loop}}

\author{G. Giavarini}
\address{{\it INFN Gruppo collegato di Parma and Dipartimento di
              Fisica dell'Uni\-ver\-sit\`a di Parma, \break
              Viale delle Scienze, 43100 Parma, Italy}}

\author{C. P. Martin}
\address{{\it Departamento de F\'\i sica Te\'orica,  C-XI,
	      Universidad Aut\'onoma de Madrid, \break
	      Cantoblanco, 28049 Madrid, Spain}}

\author{ F. Ruiz Ruiz}
\address{{\it NIKHEF-H, Postbus 41882, 1009 DB Amsterdam,
	      The Netherlands}}

\vskip 1 true cm

%Abstract
\noindent
We show that the renormalized vacuum expectation value of the Wilson
loop for topologically massive abelian gauge theory in $\RR^3$ can be
defined so that its large-mass limit be the renormalized vacuum
expectation value of the Wilson loop for abelian Chern-Simons
theory also in $\RR^3$.

\vskip 1.5 true cm

\centerline{(To appear in Nuclear Physics B)}
\endpage

\pagenumber=2

{\bf\chapter{Introduction}}

Abelian Chern-Simons theory on an oriented three-dimensional
riemannian manifold ${\cal M}$ is perhaps the simplest instance of
a topological field theory. Its classical action $S_{\ACS}$ is
given by the metric-independent integral
$$
S_{\ACS}\,=\,-\,{i\over 4\theta}\int_{\cal M} A\wedge dA \>,
\eqn\acsaction
$$
where $A$ is a $U(1)$ gauge field on ${\cal M}$.  The interest in this
theory is that it provides field theoretical definitions of
topological invariants of both the manifold and the curves lying on
it. For example, if ${\cal M}$ is compact, it is well known that the
partition function gives a topological invariant of ${\cal M}$ known
as the Ray-Singer torsion [\schwartz]. It is also well understood
that if ${\cal M}=\RR^3$, the vacuum expectation value
$$
W_{\ACS}(C) = \Bigl\langle \exp \Bigl( i \oint_C \! dx^\m A_\m
     \Bigr) \Bigr\rangle_{\ACS}
\eqn\pathint
$$
of the Wilson loop operator along a simple (\ie\ without
self-intersections) closed curve $C$ in ${\cal M}$ can be related to a
topological invariant of the curve $C$ known as the self-linking
number [\invariant] [\gauss]. Let us discuss this last point in
somewhat more detail. Using that the action $S_{\ACS}$ is quadratic in
$A$, the path integral in eq. \pathint\ can be explicity performed. In
the covariant Landau gauge $\partial A=0$ one obtains
$$
W_{\ACS}(C) = \exp\,\biggl\{
    - \,{1\over 2} \int_0^\ell\!d\tu \!\int_0^\ell\!d\td ~
    {\dot x}^\m(\tu)\>
    \bigl\langle A_\m \bigl( x(\tu) \bigr)\, A_\n \bigl( x(\td) \bigr)
    \bigr\rangle_{\ACS} \, {\dot x}^\n (\td)\, \biggr\} \> ,
\eqn\wilacs
$$
where $\ell$ and $\tau$ are respectively the length and the natural
length parameter of the curve and
$$
\VEV{A_\m(x) A_\n(y)}_{\ACS} = {i\theta\over 2\pi} \>
                   \ee_{\m\r\n} \> {(x-y)^{\r}\over |x-y|^3}
\eqn\acsprop
$$
is the gauge field propagator. Note however that the propagator
$\VEV{A_\m(x) A_\n(y)}_{\ACS}$ is not well defined at $x=y$ so the
right-hand side in eq. \wilacs\ becomes dependent on the
regularization prescription used to regularize $\VEV{A_\m(x)
A_\n(y)}_{\ACS}$ at $x\!=\!y$\foot{Rigourously speaking, the propagator
$\VEV{A_\m(x) A_\n(y)}_{\ACS}$ is a distribution of which it is known
that on any open set in $\,\RR^3\times \RR^3 - \{x\!=\!y\}\,$ is
defined by the function on the right-hand side in eq. \acsprop. The
problem is that the domain of integration in eq. \wilacs\ includes
$x=y$ and so the value of the integral depends on the definition of
$\VEV{A_\m(x) A_\n(y)}_{\ACS}$ as a distribution. But the process of
extending a function to a distribution is in general not unique and is
what in quantum field theory is called regularization. Think for
example of the distributions principal value $\,{\rm PV}(1/x)\,$ and
finite part $\,{\rm FP}(1/x)$: they are both associatd to the function
$1/x$ but as distributions are different.}.  The point is that this
prescription can be chosen so that $W_{\ACS}(C)$ yields the Gauss
self-linking number $\SL(C)$ of the curve $C$. Indeed, if a framing of
the curve, defined by a unit vector field $n^\m(\tau)$ orthogonal to
$x^\m(\tau)$, is used as a regulator, the exponent on the right-hand
side in eq. \wilacs\ is equal to $i\theta$ times the curve's
self-linking number [\gauss]:
$$
\lim_{\eta\to 0} \,\biggl\{ \, {1\over 2}
   \int_0^\ell\!d\tu \!\int_0^\ell\!d\td ~ {\dot x}^\m(\tu) \>
   \bigl\langle A_\m \bigl( x(\tu) \bigr) \,
   A_\n \bigl( x(\td) + \eta\, n(\td) \bigr) \bigr\rangle_{\ACS}\,
   \bigl( {\dot x}^\n(\td) + \eta\,{\dot n}^\n(\td) \bigr)\,
   \biggr\} \
$$
\vskip-.8truecm
$$
=\,-\,i\,\theta\,\SL(C) \>.  \qquad\qquad\qquad\qquad\qquad\qquad
               \qquad\qquad\qquad\qquad\qquad\quad
\eqn\gausslintwo
$$
As a result, $W_{\ACS}(C)$ becomes
$$
W_{\ACS}^{ren} (C) \,= \, \lim_{\eta\to 0}\, W^{reg}_{\ACS}(C) \,=\,
    e^{\,i\, \theta \,\SL(C)} \, .
\eqn\gausslin
$$
One can think of this equation as defining a renormalization scheme
in which the renormalized vacuum expectation value of the Wilson loop
along the curve $C$ gives the exponential of $i\theta$ times the
self-linking number of the curve.

It is important to stress that the value of the integral on
the right-hand side in eq. \wilacs\ depends on the regularization
chosen to define $\,{\dot x}^\m(\tu)\VEV{ A_\m \big( x(\tu) \big)\,
A_\n \big( x(\td) \big)}_{\ACS} {\dot x}^\n(\td)\,$ at
$x(\tu)=x(\td)$\foot{Note in  contrast that contributions to
$W_{\ACS}(C)$ from compact regions $|x(\tu)-x(\td)| \geq \varepsilon
> 0$ are regularization independent since the absolute value of the
propagator $\>\big\langle A\big(x(\tu)\big) \,A\big(x(\td)\big)
\big\rangle_{\ACS}\>$ is bounded in these regions.}
and that not every choice of regularization and renormalization
renders $W_{\ACS}(C)$ a topological invariant. For instance, the
regularization and renormalization used by Polyakov [\polyakov]
gives for $W_{\ACS}(C)$ the exponential of $i\theta$ times the
writhing number $w(C)$ of the curve $C$, but is well known that the
writhing number is not a topological invariant [\gauss] [\writhing].
Thus a definition of abelain Chern-Simons theory by means of
Polyakov's regularization, though useful for other purposes
[\polyakov], lacks topological meaning. In the sequel, whenever we
refer to quantum abelian Chern-Simons theory, we shall be referring
to the topologically invariant formulation in eqs. \gausslintwo\
and \gausslin.

Let us consider now a different gauge theory, namely topologically
massive abelian gauge theory. Its classical action on the manifold
${\cal M}$ endowed with metric $g_{\m\n}$ has the form [\deser]
[\TMYM]
$$
S_{\TMA} \,= \, \int_{\cal M} \bigg(
          - \,{i\over 4\theta}\, A\wedge dA \,
          + \, {1\over 16\,m}\, F \wedge * F \,\bigg) \>,
\eqn\tmaaction
$$
where $m$ is a parameter with dimensions of mass. The action now
depends on the metric through the Maxwell term $F\wedge * F$.
However, for gauge connections that grow slower than $m^{1/2}$
the metric dependence disappears in the large $m$ limit. So
starting from $S_{\TMA}$ and sending $m$ to infinity one recovers
the Chern-Simons action in eq. \acsaction. We would like to know
whether this classical convergence of topologically massive abelian
gauge theory to abelian Chern-Simons theory as $m$ goes to infinity
holds at the quantum level. More precisely, if we restrict ourselves
to ${\cal M}=\RR^3$, the question we want to address is: Is it
possible to define the renormalized vacuum expectation value of the
Wilson loop in topologically massive abelian gauge theory so that its
large $m$ limit yields the topological invariant in eq. \gausslin?
The purpose of this paper is to answer this question in the
affirmative (for an analysis of the effective action in the
non-abelian case, see ref. [\TMYM]).

The layout of the paper is as follows. In sect. 2 we introduce a
regularization method and a renormalization scheme for the vacuum
expectation value of the topologically massive Wilson loop that yields
the topological invariant in eq. \gausslin\ as $m$ goes to infinity
in a strong sense. Sect. 2 also contains our results and
conclusions, leaving for sect. 3 the details of our computations.

{\bf\chapter{Results and conclusions}}

Our concern here is the topologically massive vacuum expectation
value
$$
W_{\TMA}(C) \,=\, \Bigl\langle \exp \Bigl( i \oint_C \! dx^\m A_\m
     \Bigr) \Bigr\rangle_{\TMA}
$$
of the Wilson loop operator along a simple closed curve $C$ contained
in $\RR^3$. In the covariant Landau gauge, the latter is given by
$$
W_{\TMA}(C) \,=\, \exp\,\bigl[\,-\,F(C,m)\,\bigr] \>,
\eqn\wiltma
$$
where
$$
F(C,m) = {1\over 2} \int_0^\ell\! d\tu \int_0^\ell\! d\td ~
      \dot{x}^\m(\tu)\, \bigl\langle A_\m\bigl( x(\tu) \bigr)
      A_\n\bigl( x(\td) \bigr) \bigr\rangle_{\TMA} \dot{x}^\n(\td)
\eqn\argument
$$and
$$
\VEV{A_\m(x) A_\n(y)}_{\TMA}\,=\, 2m \idp \biggl\{
    { m \,\ee_{\m\r\n} p^\r \over p^2\,(p^2+m^2) }
    + { p^2 g_{\m\n} - p_\m p_\n \over p^2\,(p^2+m^2)} \biggr\}
    \> e^{i p\,(x-y)}
\eqn\tmaprop
$$
is the propagator of the gauge field. In deriving eqs.
\wiltma-\tmaprop\ we have rescaled $\,A\to A \theta\,$ and
$\,m \to m\theta/2\,$ in $S_{\ACS}$ and $S_{\TMA}$ so as to be rid of
the parameter $\theta$ (note that this will not affect the analysis
of the large $m$ limit since $\theta$ remains finite as $m$ goes to
infinity). Now, on every compact domain
$\left\{(x,y):|x-y| \geq \varepsilon >0 \right\}$ the propagator
$\VEV{A_\m(x) A_\n(y)}_{\TMA}$ is well defined and converges uniformly
to $\VEV{A_\m(x)A_\n(y)}_{\ACS}$ as $m$ goes to infinity. However,
when $x=y$, $\VEV{A_\m(x) A_\n(y)}_{\TMA}$ diverges linearly and the
double integral $F(C,m)$ becomes logarithmically divergent [\wpapers].
So to have a sensible definition for $W_{\TMA}(C)$ and to be able to
later evaluate its large $m$ limit, it is necessary to provide
$F(C,m)$ with a well defined meaning. This is achieved via
renormalization and entitles regularization as a first step.
Regularization can be accomplished by regularizing the integrand
$$
f(C,m;\tu,\td) \,=\, {\dot x}^\m(\tu) \>
   \bigl\langle A_\m\bigl( x(\tu) \bigr)\, A_\n \bigl(x(\td) \bigr)
   \bigr\rangle_{\TMA}\,{\dot x}^\n(\td)
\eqn\integrand
$$
in eq. \argument\ in such a way that (i) one recovers the
finite unregularized value of $f(C,m;\tu,\td)$ for
$x(\tu)\neq x(\td)$ when the regulator is removed, and (ii) no
singularity occurs for $x(\tu)=x(\td)$ when the regulator is on.

Refs. [\wpapers] propose to use a framing of the curve, defined
by a unit vector field $n^\m(\tau)$ normal to $x^\m(\tau)$, as a
regulator. Furthermore, they show that
$$
\lim_{m\to \infty} \,\lim_{\eta\to 0} \, {\rm Im}\,
   \biggl\{ {1\over 2} \int_0^\ell\! d\tu \!\int_0^\ell\!d\td ~
   {\dot x}^\m(\tu) \, \bigl\langle A_\m\bigl( x(\tu) \bigr) \,
   A_\n \bigl( x(\td) + \eta\, n(\td) \bigr) \bigr\rangle_{\TMA}
   \big({\dot x}^\n(\td) + \eta\,{\dot n}^\n(\td) \big) \!
   \biggr\}
$$
\vskip-.8truecm
$$
=\,-\,w(C) \>. \qquad\qquad\qquad\qquad\qquad\qquad
               \qquad\qquad\qquad\qquad\qquad ~
\eqn\mlimitp
$$
Let us stop for a moment and understand this result. Taking $x\neq y$,
computing the Fourier transform in eq. \tmaprop\ and retaining only
the imaginary part, one obtains [\wpapers]
$$
\Delta_{\m\n}(x-y) \equiv
   {\rm Im}\,\bigl[\,\langle A_\m(x)\,A_\n(y)\rangle_{\TMA} \,\big]
   = {i\over 2\pi}\>\ee_{\m\r\n}\>{(x-y)^\r \over |x-y|^3} \>
   \left[ \, 1 - \bigl( 1 + m\, |x-y| \bigr)\,e^{-m|x-y|}\,
   \right] \>.
$$
Using that $\Delta_{\m\n}(x-y)$ remains bounded as $x\to y$,
one concludes that the limit $\eta\to 0$ in eq. \mlimitp\ is
independent of $n^\m(\tau)$ and equal to
$$
{1\over 2} \int_0^\ell\!d\tu \!\int_0^\ell\!d\td ~ \dot{x}^\m(\tu)
   ~ \Delta_{\m\n}\big( x(\tu)-x(\td)\big) ~ \dot{x}^\n(\td) \>.
\eqn\oddpart
$$
It is important to bear in mind that this way to proceed assigns the
propagator an imaginary part at $x=y$ by first taking $x\neq y$ and
then sending $x\to y$ in the result but that this does not mean that the
imaginary part of the propagator is unambiguously defined at $x=y$,
for as already discussed the propagator itself is undetermined at
$x=y$. Finally, one shows [\wpapers] that the large $m$ limit of eq.
\oddpart\ is equal to minus the writhing number $w(C)$. Now, since
\oddpart\ is equal to the ``naive imaginary part'' of $F(C,m)$
in eqs. \wiltma-\argument, one would conclude that the large $m$ limit
of $W_{\TMA}(C)$ yields Polyakov's phase factor $e^{\,i\,w(C)}$.
Consequently, abelian Chern-Simons theory could not be obtained as the
strong large $m$ limit of topologically massive abelian gauge theory.
This is the conclusion reached in refs.  [\wpapers].

The previous analysis is however incomplete, for it overlooks that
the double integral $F(C,m)$ is logarithmically divergent and that
therefore it does not make sense to talk about its imaginary part
as being finite and equal to \oddpart, as done in refs. [\wpapers].
The divergence of $F(C,m)$ will show up as a pole in dimensional
regularization, the regularization method we will be using here.
To make rigorous statements about the large $m$ behaviour of
$W_{\TMA}(C)$, one first has to renormalize it. This is tantamount to
renormalizing the double integral $F(C,m)$. Once $F(C,m)$ has been
renormalized, it will then make sense to compute its imaginary part
and ask whether or not it leads to Polyakov's phase factor
$e^{\,i\,w(C)}$ when $m$ is sent to infinity. Of course, if the
imaginary part of the renormalized $F(C,m)$ did not depend on the
renormalization prescription, the large $m$ limit of $W_{\TMA}(C)$
would yield Polyakov's phase factor. Were this the case, the claim
made in refs. [\wpapers] would be correct and abelian Chern-Simons
theory could not be obtained from topologically massive abelian gauge
theory by taking the large $m$ limit in a strong sense.  In this
paper, we shall see however that the imaginary part of the
renormalized double integral $F(C,m)$ does depend on the
renormalization prescription and, moreover, that there is a definition
of the renormalized value of $W_{\TMA}(C)$ that converges to
$W^{ren}_{\ACS}(C)$ in eq. \gausslin\ as $m$ goes to infinity.

To renormalize the vacuum expectation value of the topologically
massive Wilson loop, we first introduce a regularization method and
then provide a renormalization scheme. Our regularization method
combines a framing of the curve and dimensional regularization in the
following way. It defines $W^{reg}_{\TMA}(C)$ as
$$
W^{reg}_{\TMA}(C) \,=\, \exp\,\bigl[\,-\,F_{reg}(C,m)\,\bigr] \>,
\eqn\regdef
$$
where
$$
F_{reg}(C,m) \,= \int_0^\ell \!d\tu \!\int_0^\ell \! d\td ~
                   f_{reg}(C,m;\tu,\td)
\eqn\regint
$$
and
$$
f_{reg}(C,m;\tu,\td) \,=\, f_1 (\tu,\td) \,+\, f_2 (\tu,\td)
                 \,+\, f_3 (\tu,\td) \>.
\eqn\regularization
$$
The functions $f_{1,2,3}(\tu,\td)$ are defined by the equations
$$
f_1 (\tu,\td) \,=\, {\dot x}^\m(\tu) \>\biggl\{ \,
    \kappa^{3-D} \idpD {\ee_{\m\r\n} p^\r \over p^2} ~
    e^{ip \big( x(\tu)-x(\td)-\eta\,n(\td) \big)} \biggr\} \>
    \big({\dot x}^\n (\td)+\eta\,{\dot n}^\n(\td) \big)
\eqn\oone
$$
$$
f_2 (\tu,\td) \,=\,  - \,{\dot x}^\m(\tu) \>\biggl\{ \,
    \kappa^{3-D} \idpD {\ee_{\m\r\n} p^\r \over p^2+m^2} ~
    e^{ip\big( x(\tu)-x(\td) \big)}\biggr\} \>
    {\dot x}^\n (\td)
\eqn\otwo
$$
$$
f_3 (\tu,\td) \,=\,  {\dot x}^\m (\tu) \> \biggl\{ \,
    m\,\kappa^{3-D} \idpD {p^2 g_{\m\n} - p_\m p_\n \over
    p^2\,(p^2+m^2)} ~ e^{ip\big( x(\tu)-x(\td) \big)} \biggr\} \>
    {\dot x}^\n (\td)   \>,
\eqn\othree
$$
where $n^\m(\tau)$ denotes the unitary vector field orthogonal to the
curve $C$ defining the framing and $\kappa$ the mass scale introduced
by dimensional regularization. Note that our regularization depends on
two regulators. One is $\eta$, the regulator governing the framing or
point-splitting in eq. \oone; the other one is $\varepsilon=D-3$, the
dimensional regulator. The regulators $\eta$ and $\varepsilon$ are not
to be removed in an arbitrary way when UV divergences arise: the
prescription is to take first the limit $\varepsilon \to 0$ and then
send $\eta$ to zero. We shall see that this, together with a suitable
renormalization scheme, gives for the large $m$ limit of the Wilson
loop the topological invariant in eq. \gausslin.

The reasons why eqs. \regint-\othree\ define a suitable
regularization of the double integral $F(C,m)$ can be explained
as follows. First, because $F_{reg}(C,m)$ defined in
eq. \regint\ above is finite for $\eta\neq 0$ and $\varepsilon$ in a
suitable domain in the complex plane. And secondly, because if
$$
U = [0,\ell]\times [0,\ell] - \{(\tu,\td):\>\tu=\td\}
  - \{(0,\ell)\} - \{(\ell,0)\}
$$
denotes the subdomain for which $x(\tu)$ is never equal to $x(\td)$
and $U^c$ is any simply connected and closed subset of $U\!$, it is
not difficult to see that the following equations hold:
$$
\lim_{\eta\to 0} \, \lim_{D\to 3} \, f_{reg}(C,m;\tu,\td) \,=\,
      f(C,m;\tu,\td) \qquad {\rm for~all}~(\tu,\td)~{\rm in}~U^c
$$
\vskip -.4 true cm
$$
\lim_{\eta\to 0} \, \lim_{D\to 3}\,
   {\int\!\!\!\int}_{\!\!\!U^c} d\tu d\td ~ f_{reg}(C,m;\tu,\td) \,=
   {\int\!\!\!\int}_{\!\!\!U^c} d\tu d\td ~ f(C,m;\tu,\td) \>.
$$
In other words, our prescription (i) provides a finite $F_{reg}(C,m)$
and (ii) does not change the value of unregularized contributions to
$F(C,m)$ as far as the latter are well defined. So in accordance with
the principles of renormalization theory [\hepp] it defines a
regularization method.

The regularized vacuum expectation value of the topologically massive
Wilson loop as computed with this regularization method turns out to
be (see sect. 3 for its derivation)
$$
W^{reg}_{\TMA}(C) \,=\, \exp \, \biggl\{ \, i\,\SL(C)
      - {m\over 2\pi}\> \biggl[ \,{1\over D-3} \,
      + \, {1\over 2}\>\ln \biggl( {m^2\over 4\pi\kappa^2} \biggr) \,
      + \, {1\over 2}\> \ga \, \Bigr] \, \ell \,
      + \, {\rm v.t.} \biggr\} \> ,
\eqn\regwilson
$$
where $\SL(C)$ is the self-linking number of the curve $C$,
$\ga$ is the Euler constant and ``${\rm v.t.}$'' stands for
contributions that vanish as $D\to 3$, $\eta\to 0$ and $m\to \infty$.
As anounced earlier, $W^{reg}_{\TMA}(C)$ becomes singular as $D$ goes
to 3. To remove this divergence, we follow refs. [\wrenormalized] and
define the renormalized vacuum expectation value of the
topologically massive abelian Wilson loop as
$$
W^{ren}_{\TMA}(C) \,=\, \lim_{\eta\to 0}\,\lim_{D\to 3}\,
   e^{\, M \ell}  ~ W^{reg}_{\TMA}(C) \>,
\eqn\renwilson
$$
where
$$
M \,= \, {m\over 2\pi}\> \biggl[ \,{1\over D-3} \,
    + \, {1\over 2}\>\ln \biggl( {m^2\over 4\pi\kappa^2} \biggr) \,
    + \, {1\over 2}\> \ga \, \Bigr] \> .
$$
As usual [\wrenormalized], $M$ can be physically interpreted as the
renormalization constant for the mass of a test particle being driven
along the the curve $C$. From eqs. \regwilson\ and \renwilson\ we
conclude that
$$
W_{\TMA}^{ren}(C) \,=\, e^{\,i\,\SL(C) \,+\, v(m)}\, ,
\eqn\resultone
$$
with $v(m)$ collecting the contributions in $``{\rm v.t.}"$
that do not vanish as $D\to 3$ and $\eta\to 0$ but do vanish as
$m\to\infty$,
$$
\lim_{m\to\infty}\,v(m) = 0\,.
\eqn\resulttwo
$$

Eq. \renwilson\ defines our renormalization scheme and eq.
\resultone\ displays the renormalized vacuum expectation value
$W_{\TMA}^{ren}(C)$ in this scheme of the topologically massive Wilson
loop. Since $W_{\TMA}^{ren}(C)$ is finite, we are now in a position to
study its large $m$ limit properly. The latter is trivial, for eqs.
\resultone\ and \resulttwo\ imply
$$
\lim_{m\to\infty} W_{\TMA}^{ren}(C) \,=\, e^{\,i\,\SL(C)} \,,
\eqn\mainresult
$$
that together with eq. \gausslin\ lead us to the main conclusion
of this paper: Abelian Chern-Simons theory can be understood as the
strong large $m$ limit of topologically massive abelian gauge theory.
This conclusion agrees with the results obtained within the
hamiltonian formalism [\hamiltonian] and entails the
definition of the renormalized topologically massive Wilson loop as
provided by eqs. \regwilson\ and \renwilson. Some final comments are
in order:

{\it Comment 1}. It is not difficult to convince oneself that if one
computes the regularized vacuum expectation value of the topologically
massive Wilson loop using the pure framing or point-splitting
regularization advocated in refs. [\wpapers], one obtains
$$
W^{\prime\,ren}_{\TMA}(C) \,=\, \lim_{\eta\to 0}\, e^{\,M'\ell}
   ~ W^{\prime\,reg}_{\TMA}(C) \,=\, e^{\,i\,w(C) \,+\, v(m)}
\eqn\renwilsont
$$
for an appropriate choice of $M'$. Here $w(C)$ denotes as usual the
writhing number of the curve $C$. It then follows that
$$
\lim_{m\to\infty} W^{\prime\,ren}_{\TMA} \,=\, e^{\,i\,w(C)}\, .
\eqn\largemassw
$$
Eqs. \renwilsont\ and \largemassw\ tell us the way the results
presented in refs. [\wpapers] should be understood, namely as
corresponding to a particular renormalization scheme defined from a
particular regularization method. The conclusion in refs. [\wpapers]
should therefore be restated to read that for the particular choice
of renormalization scheme \renwilsont\ the large $m$ limit of the
renormalized Wilson loop is not a topological invariant. Our
renormalization scheme though yields a topologically invariant large
$m$ limit.

{\it Comment 2}. According to general principles of renormalized
quantum field theory [\wrenormalized] [\epstein], since
$W_{\TMA}^{ren}(C)$ and $W^{\prime\,ren}_{\TMA}(C)$ in eqs.
\resultone\ and \renwilsont\ correspond to two different
renormalization schemes, there should be a local
parametrization-invariant counterterm that transforms
$W_{\TMA}^{ren}(C)$ into $W^{\prime\,ren}_{\TMA}(C)$. Such a
counterterm does indeed exist and has the form
$$
e^{\,i\, \TW(C)} \, ,
$$
where $\TW(C)$ is the functional
$$
\TW(C) \,=\, {1\over 2\pi} \int_0^\ell \!d\tau \>
   \ee_{\m\r\n} \,{\dot n}^\n(\tau)\, {\dot x}^\r(\tau)\,
   n^\n(\tau)  \>.
\eqn\torsion
$$
To see the latter, it is enough to multiply
$W^{\prime\,ren}_{\TMA}(C)$ by $e^{\,i\,\TW(C)}$ and use the identity
$\SL(C)=w(C)+\TW(C)$ [\writhing] since then
$$
W_{\TMA}^{ren}(C) \,=\, e^{\,i\,\TW(C)} ~
          W^{\prime\,ren}_{\TMA}(C) \>.
$$
The number $\TW(C)$ is the twist of the ribbon determined by the
curve $C$ and its framing $n^\m(\tau)$ and is not a topological
invariant [\writhing]. Notice that $e^{\,i\,\TW(C)}$ is a
perfectly allowed counterterm since $\TW(C)$ is a local
integrated functional of $\,x^\n(\tau),~\,n^\n(\tau)\,$ and its
derivatives and is parametrization invariant. Note also that $\TW(C)$
has support on the compact segment line $\{\,(\tu,\td):\>\tu=\td\}$
only. This can be easily understood if one takes into account that
contributions to $F(C,m)$ from non-coincident points
$x(\tu)\neq x(\td)$ are regularization independent and do not enter
in neither of the renormalizations performed in eqs. \renwilson\
and \renwilsont, so any difference between $W_{\TMA}^{ren}(C)$
and $W^{\prime\,ren}_{\TMA}(C)$ has to be of the form the
exponential of the integral of a function with support only on the
compact segment line $\{\,(\tu,\td):\>\tu=\td\}$. To summarize, once a
topologically massive renormalized Wilson loop has been obtained along
the lines discussed in refs. [\wpapers], a finite renormalization (or
change of renormalization scheme) by means of the counterterm
$e^{\,i\,\TW(C)}$ leads to a renormalized Wilson loop whose large $m$
limit is a topological invariant. This is the same topological
invariant that our renormalization scheme yields automatically.

\chapter {\bf Computations}

In this section we give some details of the derivation of the
regularized vacuum expectation value of the topologically massive
Wilson loop in eq. \regwilson. We shall prove the three
following partial results:
$$
F_1(C) \,\equiv \int_0^\ell \!d\tu \! \int_0^\ell \!d\td \>
    f_1(\tu,\td) \,=\, -\,i\, \SL(C) \,+\, {\rm v.t.}
\eqn\resultoone
$$
\vskip -.4true cm
$$
F_2(C,m) \,\equiv \int_0^\ell \!d\tu \! \int_0^\ell \!d\td \>
    f_2(\tu,\td) \,=\, 0 \,+\, {\rm v.t.}
\eqn\resultotwo
$$
and
$$
F_3(C,m) \,\equiv \int_0^\ell \!d\tu \! \int_0^\ell \!d\td \>
    f_3(\tu,\td) \,=\, {m\over 2\pi}\, \biggl[ \,{1\over D-3} \,
    + \, {1\over 2}\>\ln \biggl( {m^2\over 4\pi\kappa^2} \biggr) \,
    + \, {1\over 2}\> \ga \, \biggr] \, \ell \,+\, {\rm v.t.} ~,
\eqn\resultothree
$$
where ``${\rm v.t.}$'' stands for contributions that vanish when the
limits $D\to 3,~\eta\to 0$ and $m\to \infty$ are taken in this order.
The three equations \resultoone-\resultothree, together with
$$
F_{reg}(C,m) \,=\, F_1(C) \,+\, F_2(C,m) \,+\, F_3(C,m)
$$
and eq. \regdef,  imply eq. \regwilson. So to obtain eq. \regwilson\
all we have to do is derive eqs. \resultoone-\resultothree.

\subsection{Derivation of eq. \resultoone}

To perform the integral over the momenta in $f_1(\tu,\td)$ in eq.
\oone\ we use
$$
\idpD \> {e^{ipx} \over p^2} \,=\>
    { \gm\!\left( {D\over 2}-1 \right) \over 4\,\pi^{D/2}} ~
    {1 \over |x|^{D-2}} \>.
$$
This gives
$$
\eqalign{
F_1(C) \,= & \, {i\> \kappa^{3-D} \> \gm(D/2) \over 4\,\pi^{D/2}}  \cr
{\scriptstyle \times} & \int_0^\ell \!d\tu \!\int_0^\ell \!d\td
      ~ \ee_{\m\r\n}\> {\dot x}^\m(\tu) ~
      {x^\r(\tu)-x^\r(\td)-\eta\, n^\r(\td) \over
                          |x(\tu)-x(\td)-\eta\, n(\td)|^D} ~
      \bigl( {\dot x}^\n(\td)+\eta {\dot n}^\n(\td) \bigr) \>.\cr}
$$
Noting now that the integral on the right-hand side is absolutely
convergent at $D=3$, we have that the following equation holds in
dimensional regularization [\collins]:
$$
\lim_{D\to 3}\, F_1(C) \,=\, {i\over 4\pi} \int_0^\ell \!d\tu
   \!\int_0^\ell \!d\td ~ \ee_{\m\r\n} \> {\dot x}^\m(\tu) \>
   {x^\r(\tu)-x^\r(\td)-\eta\, n^\r(\td) \over
                     |x(\tu)-x(\td)-\eta\, n(\td)|^3} \>
   \bigl( {\dot x}^\n(\td)+\eta {\dot n}^\n(\td) \bigr) \>.
$$
Taking finally the limit $\eta\to 0$ and recalling eqs. \acsprop\ and
\gausslintwo, we conclude
$$
\lim_{\eta\to 0}\,\lim_{D\to 3}\, F_1(C) \,=\, -\,i\, \SL(C) \>.
$$

\subsection{Derivation of eq. \resultotwo}

In the remainder of this section we show that eqs. \resultotwo\ and
\resultothree\ hold true for any analytic curve $x^\m(\tau)$. As will
become clear in a moment, analiticity is a technical requirement that
will render uniformly convergent the series expansions we perform
below.  The convergence being uniform will make possible to integrate
these series and to evaluate their large $m$ limit term by term. Our
computations will then be strictly valid for analytic curves only,
though we believe our results are correct for curves of class $C^1$.
In this case however some other method should be employed to prove
them.

Calling $R(\tau)$ to the radius of convergence of $x^\m(\tau)$ and
using that by assumption $R(\tau)$ is positive for all $\tau$, we
can define $\d$ such that
$$
0 < \d < {\rm min}\,\{R(\tau): \> 0\leq \tau \leq \ell\,\} \>.
\eqn\radius
$$
So writing $F_2(C,m)$ as
$$
F_2(C,m) \,=\,2 \int_0^\ell\!d\tu \! \int_0^{\tu} \!d\td
    ~ f_2(\tu,\td)
$$
and splitting its domain of integration
$$ {\cal D}=\{\,(\tu,\td):~ 0 \leq \tu \leq \ell\,,
                        ~\, 0 \leq \td \leq \tu \,\}
$$
into three subdomains
$$
\eqalign{
& {\cal D}_1 = \{\,(\tu,\td):~ 0\leq \tu \leq \ell\,,
                           ~\, 0 \leq \tu-\td \leq \d \,\} \cr
& {\cal D}_2 = \{\,(\tu,\td):~ \ell-\d \leq \tu \leq \ell\,,
                           ~\, 0 \leq \td \leq \tu - \ell+ \d \,\} \cr
& {\cal D}_3 = {\cal D} - {\cal D}_1 - {\cal D}_2 \>,  \cr}
\eqn\domains
$$
we have
$$
F_2(C,m) \,=\, \sum_{i=1}^3 F_{2,{\cal D}_i} \> ,
\eqn\decom
$$
where
$$
F_{2,{\cal D}_i} \,=\, 2 {\int\!\!\!\int}_{\!\!\!{\cal D}_i}
     \!d\tu\,d\td \> f_2(\tu,\td) \> .
$$
Using
$$
\idpD \> {e^{ipx}\over p^2+m^2} \,=\, {1\over (4\pi)^{D/2}}\>
    \int_0^\infty {d\la \over \la^{D/2}} ~
    e^{-\la m^2}\>e^{-x^2/4\la}
\eqn\schwinger
$$
for the integration over the momenta hidden in $f_2(\tu,\td)$, we
obtain
$$
\eqalign{
F_{2,{\cal D}_i} \,=\, -\, {i\, \kappa^{3-D}\over (4\pi)^{D/2}} \,
    {\int\!\!\!\int}_{\!\!\!{\cal D}_i} \!d\tu\,d\td \>
    \!\int_0^\infty & {d\la \over \la^{(D+2)/2}} ~
    \ee_{\m\r\n} ~ {\dot x}^\m(\tu) \>
    \big( x^\r(\tu)- x^\r(\td) \big) \> {\dot x}^\n(\td) \cr
& {\scriptstyle \times} ~ e^{-\la m^2} \> e^{-(x(\tu)-x(\td))^2/ 4\la}
    \,. \cr}
\eqn\intodd
$$
In the sequel we consider $F_{2,{\cal D}_i}$ for each $i$ separately
and show that
$$
\lim_{m\to\infty}\,\lim_{D\to 3}\, F_{2,{\cal D}_i} \,=\, 0
    \qquad i=1,2,3.
\eqn\odd
$$
This result, along with eq. \decom, implies eq. \resultotwo.

We start proving eq. \odd\ for $\,i=1$. Performing a change of
variables $(\tu,\td)\,\to\, (\tu,t)$, with $t=\tu-\td$, the domain
of integration becomes
$$
{\cal D}_1 = \{\,(\tu,t):~ 0\leq \tu \leq \ell\,,
                       ~\, 0 \leq t \leq \d \,\} \>.
$$
Since $x^\m(\tau)$ is analytic with radius of convergence $R(\tau)$
larger than $\d$ for all $\tau$, $x^\m(\td)$ admits a uniformly
convergent power series on ${\cal D}_1$:
$$
x^\m(\td) \,=\, x^\m(\tu-t) \,=\, \sum_{n=0}^\infty \>
     \left[\,{d^n \over d\tu^n} \> x^\m(\tu)\,\right] \>
     {(-t)^n\over n!} \>.
\eqn\expansion
$$
Substituting eq. \expansion\ in eq. \intodd\ and using that
uniform convergence implies that integration over $t$ can be
performed term by term, we have
$$
F_{2,{\cal D}_1} \,=\, - \> {i\,\kappa^{3-D} \over (4\pi)^{D/2}} \>
    \int_0^\ell \!d\tu \> \sum_{j=0}^\infty \> I_j(D,m,\tu,\d) \>,
\eqn\intone
$$
where $I_j(D,m,\tu,\d)$ is given by
$$
I_j(D,m,\tu,\d) \,= \sum_{k=0}^{[j/4]} \> c_{jk}(\tu) \int_0^\d \!dt ~
   t^{4+j} \> \int_0^\infty {d\la \over \la^{(D+2+2k)/2}} ~
   e^{-\la m^2} \> e^{-t^2/4\la}  \>.
\eqn\intj
$$
Here $[x]$ denotes the integer part of $x$ and $c_{jk}(\tu)$ are
functions of $x^\m(\tu)$ and its derivatives that do not depend on $m$
nor $D$ and whose explicit form is not important. In obtaining eqs.
\intone\ and \intj\ we have used that, being $\tau$ the natural length
parameter, $\,|\,{\dot x}(\tau)\,|=1$ and $\,{\dot x}^\m(\tau) \,
{\ddot x}_\m(\tau)=0$. Integrating over $\la$ in eq. \intj\ we obtain
$$
I_j(D,m,\tu,\d) \,=\, 2\, \sum_{k=0}^{[j/4]} \> c_{jk}(\tu) \,
   (2m)^{{D\over 2}\,+\,k} \> \int_0^\d \!dt ~ t^{4+j-k-{D\over2}}
   ~ K_{{D\over 2}+k} (mt) \> ,
\eqn\bessel
$$
with $K_\a(z)$ the modified Bessel function of third type and order
$\a$. Using that for half-integer order $K_\a(z)$ takes the simple
form
$$
K_{n+{1\over 2}}(z) \,=\,\sqrt{{\pi\over 2z}} \> e^{-z} ~
    \sum_{p=0}^n ~ {(n+p)! \over p!\,(n-p)!} \> {1\over (2z)^p}
    \qquad n=0,1,2,... ,
\eqn\besselhalf
$$
it is straightforward to see that the integral over $t$ in eq.
\bessel\ is absolutely convergent at $D=3$ for all $m$ (convergence
at $t=0$ does not pose any problem). This implies that the limit
$\,D\to 3\,$ of $I_j(D,m,\tu,\d)$ can be calculated by taking $D=3\,$
inside the integral in eq. \bessel. Thus, after a change of variables
$\,u=mt$, we have
$$
\eqalign{
\lim_{D\to 3} \, I_j(D,m,\tu,\d) \, &=\, 4\sqrt{\pi}
   \, \sum_{k=0}^j \> {2^k\, c_{jk}(\tu) \over m^{2+j-2k}} \cr
&\,{\scriptstyle \times} \> \sum_{p=0}^{k+1} ~
   {(k+1+p)! \over 2^p\,p!\,(k+1-p)!} \,
   \int_0^{m\d}\! du ~ u^{2+j-k-p} ~ e^{-u} \>.
   \cr}
$$
The right-hand side obviously goes to zero as $m$ goes to infinity,
hence
$$
\lim_{m\to\infty}\,\lim_{D\to 3} I_j(D,m,\tu,\d) \,=\, 0 \>.
$$
Together with uniform convergence of the series in eq. \intone\ this
implies
$$
\lim_{m\to\infty}\,\lim_{D\to 3} F_{2,{\cal D}_1} = 0 \>,
$$
in accordance with eq. \odd. Using similar methods it is not difficult
to prove that eq. \odd\ holds for $i=2$. Lastly we consider $i=3$.
Noting that for all $(\tu,\td)$ in ${\cal D}_3$ there exists a
real number $r_0>0$ such that $|x(\tu)-x(\td)| \geq r_0$, one may
readily see that
$$
0 \leq \big\vert \, F_{2,{\cal D}_3} \, \big\vert \, \leq \,
    A_0(\d)\, \kappa^{3-D}\, {\bigg( {m\over 2\pi r_0} \bigg)}^{\!D/2}
    ~ K_{{D\over 2}}(mr_0) \>,
$$
with $A_0(\d)$ a positive constant that does not depend on $m$
nor $D$. So after taking the limits $D\to3$ and $m\to\infty$ and
using eq. \besselhalf, we conclude
$$
\lim_{m\to\infty}\,\lim_{D\to 3} F_{2,{\cal D}_3} = 0 \>.
$$
This completes the proof of eq. \odd, hence of eq. \resultotwo.

\subsection{Derivation of eq. \resultothree}

To prove eq. \resultothree\ we proceed as follows. First we realize
that the contribution to $F_3(C,m)$ of the term $p_\m p_\n$ in
$f_3(\tu,\td)$ is the integral of a total derivative along a closed
curve, so it vanishes. This leaves us with
$$
F_3(C,m) \,=\, {m\,{\kappa}^{3-D}\over (4\pi)^{D/2}} \>
    \int_0^\ell \!d\tu \!\int_0^\ell \!d\td \> {\dot x}^\m(\tu) \,
    {\dot x}_\m(\td) \> \int_0^\infty {d\la \over \la^{D/2}} ~
    e^{-\la m^2} \> e^{-\,(\,x(\tu)-x(\td)\,)^2/4\lambda} \> ,
\eqn\inteven
$$
where eq. \schwinger\ has been used for the integration over
the momenta in $f_3(\tu,\td)$. Next we write
$$
F_3(C,m) \,=\, \sum_{i=1}^3 F_{3,{\cal D}_i} \> ,
\eqn\decompthree
$$
with the domains ${\cal D}_i$ as in eq. \domains\ and
$$
F_{3,{\cal D}_i} \,=\, {2\, m\,{\kappa}^{3-D}\over (4\pi)^{D/2}}
    {\int\!\!\!\int}_{\!\!\!{\cal D}_i} d\tu \, d\td ~
    {\dot x}^\m(\tu) \, {\dot x}_\m(\td) \> \int_0^\infty
    {d\la \over \la^{D/2}} ~ e^{-\la m^2} ~
    e^{-\,(\,x(\tu)-x(\td)\,)^2/4\lambda} \>.
$$
Finally, we analyze each one the integrals $F_{3,{\cal D}_i}$.

We start with $F_{3,{\cal D}_1}$. Analogous arguments to those
used for $F_{2,{\cal D}_1}$ give
$$
\eqalign{
F_{3,{\cal D}_1} \,&=\, -\, {2m\,{\kappa}^{3-D}\over (4\pi)^{D/2}} ~\,
    \bigg\{\, \ell \int_0^\d \!dt\, \int_0^\infty
    {d\la \over \la^{D/2}} ~ e^{-\la m^2} ~ e^{-t^2/4\lambda}  \cr
&+\int_0^\ell \!d\tu \> \sum_{j=2}^\infty \>
    \sum_{k=0}^{[j/4]} \> \tilde{c}_{jk}(\tu) \int_0^\d \!dt \> t^j
    \int_0^\infty {d\la \over \la^{(D+2k)/2}} ~ e^{-\la m^2} \>
    e^{-\,t^2/4\lambda} \, \biggr\} \>, \cr}
\eqn\inttwo
$$
where $\tilde{c}_{jk}(\tu)$ are functions of $x^\m(\tu)$ and its
derivatives only. The first term on the right-hand side in eq.
\inttwo\ is not well defined at $D=3$ and thus one has to keep $D$
arbitrary. Regarding the second term, the very same arguments as those
used for eq. \intone\ show that it vanishes as $D\to 3$, $m\to\infty$.
Hence, only the first term contributes to the double limit $D\to 3$,
$m\to\infty$:
$$
F_{3,{\cal D}_1} \,=\, -\,{2\, m\,{\kappa}^{3-D}\,\ell
                                          \over (4\pi)^{D/2}} \,
    \int_0^\d \!dt \int_0^\infty {d\la \over \la^{D/2}} ~
    e^{-\la m^2} ~ e^{-\,t^2/4\lambda} \,+\, {\rm v.t.}
$$
Interchanging the order of integration over $t$ and $\la$ and making
the changes $\,u=mt\,$ and $\zeta=\la m^2$, we
have
$$
F_{3,{\cal D}_1} \,=\,-\, {m\ell\over 4\pi}\,
   {\left( {m^2\over4\pi\kappa^2} \right)}^{\!(D-3)/2}
   \int_0^\infty {d\zeta \over \zeta^{(D-1)/2}} ~ e^{-\zeta}
   ~ \Phi\!\left( {m\d \over 2\sqrt{\zeta}} \right) \> ,
$$
where
$$
\Phi(x) \,= {2\over \sqrt{\pi}} \int_0^x du \> e^{-u^2}
$$
is the probability integral. Noting that
$$
\zeta^{(1-D)/2} \,=\, {2\over 3-D} \> {d\over d\zeta}
    \> \zeta^{(3-D)/2}  \>,
$$
integrating by parts and expanding
${\big( 4\pi\kappa^2m^{-2}\zeta \big)}^{(3-D)/2}$ in powers of
$D-3$, we obtain
$$
\eqalign{
F_{3,{\cal D}_1} \,&=\, {m\ell\over 2\pi} \> {1\over D-3} \,
    +\, {m\ell\over 4\pi} \>
               \ln\!\left( {m^2\over 4\pi\kappa^2} \right) \cr
& + \,{m\ell\over 4\pi} \int_0^\infty d\zeta \> \ln\zeta ~\,
    {d\over d\zeta} \left[ \> e^{-\zeta} ~
    \Phi\!\left( {m\d \over 2 \sqrt{\zeta}} \right)\,\right] \,
    +\, O\,(D-3) \>. \cr}
$$
The first term on the right-hand side gives a pole in $D-3$, whereas
the second and third terms give finite non-vanishing contributions.
It can be seen after some calculus that the large $m$ limit of the
integral in the third term is given by
$$
m \int_0^\infty d\zeta \> \ln\zeta ~\, {d\over d\zeta}
  \left[ \, e^{-\zeta} ~ \Phi\!\left( {m\d
                          \over 2 \sqrt{\zeta}} \right)\, \right]
  \,=\, m\ga \,+\, {\rm terms~that~vanish~as~}m~{\rm goes~to}~\infty
  \>,
$$
where $\ga$ is the Euler constant. We then conclude that
$$
F_{3,{\cal D}_1}  \,=\,
    {m\over 2\pi}\> \biggl[ \,{1\over D-3} \,
    + \, {1\over 2}\>\ln \biggl( {m^2\over 4\pi\kappa^2} \biggr) \,
    + \, {1\over 2}\> \ga \, \biggr] \, \ell + {\rm v.t.}
\eqn\evenone
$$
Using the same type of arguments it is not difficult to see, though
a bit tedious, that
$$
\lim_{m\to\infty}\,\lim_{D\to 3}\, F_{3,{\cal D}_2} \,=\, 0 \>.
\eqn\eventwo
$$
Concerning $F_{3,{\cal D}_3}$, it can be studied in the same way as
$F_{2,{\cal D}_3}$. Indeed, for all $(\tu,\td)$ in ${\cal D}_3$ we
have $|x(\tu)-x(\td)| \geq r_0> 0$ and
$$
0 \leq \big\vert \, F_{3,{\cal D}_3} \, \big\vert \, \leq \,
    B_0(\d)\, \kappa^{3-D}\,
    {\bigg( {m\over 2\pi r_0} \bigg)}^{\!{D\over 2}-1}
    ~ K_{{D\over 2}-1}(mr_0) \>,
$$
with $B_0(\d)$ a positive constant. So taking the limits $D\to 3$ and
$m\to\infty$ and using eq. \besselhalf\ we have
$$
\lim_{m\to\infty}\,\lim_{D\to 3} F_{3,{\cal D}_3} = 0 \>.
\eqn\eventhree
$$
Putting together eqs. \decompthree\ and \evenone-\eventhree\ we
reach eq. \resultothree. With this we complete the proof of eqs.
\resultoone-\resultothree, hence of eq. \regwilson.

\bigskip

\noindent{\bf Acknowledgements:} \noindent FRR was supported by FOM,
The Netherlands. The authors are also grateful to CICyT, Spain for
partial support.

\refoutlw

\end